# Extreme Terahertz Nonlinearity of AlGaN/GaN-based Grating-Gate Plasmonic Crystals

*Pavlo Sai[1,2], Vadym V. Korotyeyev[1,3], Dmytro B. But[1,2], Maksym Dub[1,2], Dmitriy Yavorskiy[1,2], Jerzy Łusakowski[4], Mateusz Słowikowski[2], Serhii Kukhtaruk[1,3], Yurii Liashchuk[1,3], Jeong Woo Han[5,6], Christoph Böttger[5], Alexej Pashkin[7], Stephan Winnerl[7], Wojciech Knap[1,2], and Martin Mittendorff[5]\**

[1]Institute of High Pressure Physics PAS, 01-142 Warsaw, Poland
[2]CENTERA, CEZAMAT, Warsaw University of Technology, 02-822 Warsaw, Poland
[3]V. Ye. Lashkaryov Institute of Semiconductor Physics (ISP), NASU, 03028 Kyiv, Ukraine
[4]Faculty of Physics, University of Warsaw, ul. Pasteura 5, 02-093 Warsaw, Poland
[5]Universität Duisburg-Essen, Fakultät für Physik, 47057 Duisburg, Germany
[6]now at Department of Physics Education, Chonnam National University, Gwangju 61186, South Korea
[7]Helmholtz-Zentrum Dresden-Rossendorf, 01328 Dresden, Germany
E-mail: martin.mittendorff@uni-due.de

**Abstract**:

We present a novel approach to enhance THz nonlinearity by the resonant excitation of two-dimensional plasmons in grating-gate plasmonic crystals. Using a high-electric-field THz pump-THz probe technique, we investigate the nonlinear interaction of spectrally narrow THz pulses with plasmon oscillations in a two-dimensional electron gas on an AlGaN/GaN interface integrated with metallic grating. Nonlinear effects are observed as ultrafast, pump-induced changes in THz transmission, with relative transparency strongly dependent on plasmonic mode excitation and saturating at pump fluences of about 200 nJ cm$^{-2}$. The maximal relative transparency, reaching 45 % at 350 nJ cm$^{-2}$, occurs under resonant excitation of a localized plasmon mode at the strong electrostatic modulation of 2DEG concentration. Transient dynamics reveal ultrafast relaxation times of 15-20 ps, while the effects can be observed at elevated temperatures of up to 150 K. A nonlinear model of plasmonic crystal, based on finite-difference time-domain electrodynamic simulations coupled with viscous hydrodynamic electron transport model, elucidates key nonlinear mechanisms, including near-field effects under metallic gratings, electron heating, plasmon resonance broadening, and redshift. These results demonstrate that even conventional semiconductors such as AlGaN/GaN can achieve nonlinear THz responses comparable to or exceeding those of graphene, showing strong potential for ultrafast THz modulation and nonlinear photonics applications.

1. **Introduction**

Nonlinear phenomena are of key importance in modern photonics, facilitating a range of optical functionalities such as wavelength conversion,[1] high-order harmonic generation,[2] saturable absorption,[3] and ultrafast switching.[4, 5] Traditionally, the observed nonlinear optical effects have been predominantly weak and primarily studied in the visible and near-infrared spectral regions. However, extending these phenomena into the technologically relevant terahertz (THz) range is crucial, e.g., for advancing future 6G communication technologies, which demand deep and fast modulation of the THz carrier frequency beams. In contrast to optical nonlinearities in the visible regime that are most commonly based on a nonlinear polarization, e.g. due to the Kerr effect, also thermal effects play an important role in the THz regime. If the temperature modulation is fast enough, i.e, on the time scale of the electric field oscillation, even high harmonic generation is feasible.[6]

There are two main approaches to increase THz nonlinearity: the application of highly nonlinear materials such as graphene, which is known as a leading nonlinear material in the THz domain due to its linear, Dirac-type band dispersion,[6, 7] where its nonlinearity can be dynamically tuned via electrostatic gating;[8] and the nanoengineering of sophisticated semiconductor-based structures.[9, 10] Nanotechnology allows the fabrication of hybrid metamaterials to enhance light-matter interaction, for example, by utilizing interference effects from reflective surfaces near absorbing layers.[11, 12] Additionally, plasmonic structures with sub-micrometer resolution of resonant cells host localized two-dimensional (2D) plasmons and offer further control over nonlinear phenomena by tailoring the geometry of the plasmonic cavity or the concentration of charge carriers.[13]

The interaction of light with plasmons is particularly valuable in the THz region, enabling light confinement in deeply subwavelength spatial areas. This confinement amplifies light-matter interactions, serving as an efficient mechanism for generating nonlinear optical effects. The essential modification of plasmonic resonant band (spectral position or its width) depending on the intensity of incident THz waves is the key mechanism underlying in the discussed nonlinearity. Latter can lead to considerable changes of optical transmission characteristics of plasmonic structures including graphene microribbons [14, 15] and discs,[16, 17] achieving up to 13% pump-induced changes in THz transmission at plasmon resonance frequencies, with ultrafast relaxation times of approximately 10 ps.

In this work, we explore a novel approach to enhance THz nonlinearity by the resonant excitation of plasmons in grating-gate plasmonic crystals (PCs). These structures consist of a periodic arrangement of gated and ungated regions within a two-dimensional (2D) conducting layer integrated with a metallic grating.[18] Grating-gate PCs have become a prominent platform for investigating 2D plasmons, particularly when fabricated on high-mobility $A^{III}B^{V}$ semiconductor heterostructures with 2D electron gas (2DEG)[19-21] or graphene.[22, 23] Linear THz spectroscopy techniques such as Fourier transform far-infrared (FTIR) and THz time-domain spectroscopies (THz TDS) have demonstrated their ability to efficiently excite multiple plasmon modes.[19, 24-28] Previous studies also highlighted the enhanced performance of such structures for THz detection [29] and thermal plasmonic emission,[30-32] reinforcing their potential in this field.

Recently, delocalized and localized PC phases were identified in regimes of weak and strong modulation of 2DEG concentration, respectively.[33] This electrical plasmonic mode switching by modulating the 2DEG concentration profile using a gate voltage applied to the grating electrode. Our approach builds on these findings and offers three key advantages: (i) enhanced THz nonlinearity: the electric field is concentrated near the grating electrode, particularly in the localized phase of the PC, when plasma oscillations are localized in the ungated 2DEG regions, amplifying nonlinear interactions; (ii) tunable THz nonlinearity: the nonlinear optical response can be dynamically switched on or off by adjusting the gate voltage to align with plasmon resonance frequencies; (iii) applicability to the conventional semiconductors, benefiting from their mature and highly developed fabrication technologies. To validate this concept, we focus on an AlGaN/GaN-based semiconductor heterostructure, which is traditionally not considered to be a highly nonlinear material due to its classical parabolic band dispersion. However, GaN-based structures exhibit several advantages for linear THz plasmonics, including high electron concentrations ($\sim 10^{13}$ cm$^{-2}$), high electron mobility (up to $10^4$ cm$^2$ V$^{-1}$ s$^{-1}$ at 70 K and $2\times 10^3$ cm$^2$ V$^{-1}$ s$^{-1}$ at 300 K), and optical phonon energy exceeding 92 meV ($\sim$23 THz)[34]; these provide excellent conditions for the observation of 2D plasmon resonances up to room temperatures. By combining these material properties with the grating-gate PC design, we demonstrate that even conventional semiconductors can achieve significant nonlinear optical responses in the THz range.

To study the THz nonlinearity in AlGaN/GaN-based grating-gate PCs experimentally, we used a high electric field THz pump – THz probe spectroscopy system based on the Free Electron Laser (FEL) at the FELBE facility, Helmholtz-Zentrum Dresden-Rossendorf[35] as a tunable source of THz radiation. The FEL generates intense THz multi-cycled pulses with a duration of $\sim$10 ps and a moderate peak electric field of several kV cm$^{-1}$. The FEL-based system operates with spectrally narrow pulses with a linewidth < 0.1 THz, much narrower than spectral bands of typical plasmon resonances in AlGaN/GaN. Under resonance conditions, a significant part of the electromagnetic (*em*) energy of the pump pulses efficiently transfers to plasmonic oscillations of the 2DEG. This provides an excellent platform for exploring the nonlinear responses of PC structures associated with pump-induced transient changes of transmission spectra.

We studied two different PC samples with different grating geometries that host delocalized and localized plasmonic phases. The FEL was tuned to allow effective pumping of different plasmonic resonances by varying the gate voltage only. We observed an extremely nonlinear response as the THz pump pulse induces essential changes of THz transmission. The nonlinear effects are stronger in the localized phase of the PC structure when plasma oscillations are confined in the ungated regions of the 2DEG. Here, the pump-induced change in transmission reaches high values of 45%, accompanied by ultrafast relaxation times of approximately 15 ps. Importantly, this effect is achieved at moderate pumping fluencies not exceeding 350 nJ cm$^{-2}$, with a maximum electric field of the pump pulse at $E_{max}$ < 5 kV cm$^{-1}$, showing the potential of these findings for the future development of ultrafast THz modulators or switches.

Additionally, states with increased (positive relative transparency) or decreased (negative relative transparency) pump-induced transmittance of the PC structure were identified. The emergence of these states can be attributed to both a transient broadening and spectral redshift of the 2D plasmon resonance. The former can result from the heating of the 2DEG during the pump pulse with the activation of an additional scattering mechanism. A redshift of 2D plasmon resonance as well as the cyclotron resonance (CR) with elevating equilibrium temperature has been observed in many THz spectroscopy and THz magneto-spectroscopy measurements.[27, 36, 37] As in earlier studies, we found that the effect is related to the temperature-dependent electron effective mass. Our studies of CR also confirmed a significant increase of nearly 30% of the electron effective mass in the bare AlGaN/GaN sample in a wide temperature range of 10 – 300 K (see Section 1 of the Supporting Information).

The quantitative analysis of the experimental data was done within the framework of a developed nonlinear theory of the interaction of *em* pulses with a 2DEG under the grating. The existing theoretical approaches for simulating the optical characteristics of different kinds of grating-based PC structures are applicable only in the linear regime.[33, 36, 38-45] These papers provide the linear theories of the interaction of electromagnetic waves with 2D plasmons under the grating and most of them use the Drude-Lorentz model to relate electric field and conduction current induced in the plane of a 2DEG.

Our theoretical model of a nonlinear plasmonic crystal is based on coupled solutions of Maxwell's equations and electron nonlinear transport equations, employing a viscous hydrodynamic (HD) model. Our transport model describes and takes into account several important factors including: (i) any level of degeneracy in the 2DEG and any lateral profile of electron concentration, $n_0$; (ii) nonlinear field dependences of the electron drift velocity $V_d(E^{2D})$ and electron temperature $T_e(E^{2D})$; (iii) non-elastic electron-polar optical phonon scattering mechanism (electron-LO scattering); (iv) electron temperature dependences of viscosity and heat conductivity coefficients. The above-mentioned temperature dependence of the electron effective mass was also taken into account. It should be noted that the THz pulse, acting on the 2DEG under the metallic grating, induces strong transient and spatial modifications of all transport characteristics of the 2DEG due to the near-field effects and effects related to the viscosity behavior of the 2D electron fluid.

Thus, accurate modeling of the nonlinear effects arising from the interaction of high-field pulses with PC structures presents a considerable challenge. This complexity necessitates numerical spatio-temporal simulations of high-field pulse transmission through the PC structure, accounting for strong near-field effects (due to the adjacent metallic grating) and nonlinear electron transport models that extend beyond the conventional Drude-Lorentz model.

The paper is organized as follows: Section 2 describes the samples under investigation and provides details of the THz pump-probe spectroscopy measurements. Section 3 presents the theoretical grounds of our study, a detailed analysis of the measurement results, and their comparison with theoretical simulations. Finally, Section 4 contains further discussion and conclusions of our work.

## 2. AlGaN/GaN-based grating-gate plasmonic crystals and THz pump-probe spectroscopy measurements

The samples investigated in this study are based on AlGaN/GaN heterostructures with a carrier density of about $10^{13}$ cm$^{-2}$ and a high electron mobility of about 7500 cm$^2$ V$^{-1}$ s$^{-1}$ at 10 K. The heterostructure was grown by a metal-organic vapor phase epitaxy on a semi-insulating SiC substrate. An electron-beam lithography was used to fabricate the grating-gate structure on the sample surface, details on the sample fabrication can be found in Ref. [33]. Two samples with different grating-gate geometry were fabricated: sample S13 with a grating periodicity, $a_g$, of 1 μm and a filling factor, $f$, of 50%, and sample S8 with a periodicity of 2.5 μm and a filling factor of 72% (see insets in **Figure 1** (a) and (b)).

The FTIR transmission spectra, $T_\omega$, of both samples for two selected values of gate voltages, $V_G$, are shown in Figure 1 (a) and (b). As can be seen, we observe the fundamental plasmon resonance of the localized phase at around 1.8 THz and 1.9 THz at negative gate voltages of -3.5 V and -4 V for samples S13 and S8, respectively. At this gate voltage, the carrier density under the gate fingers is nearly completely depleted. Upon increasing the negative voltage, these localized plasmon modes are shifted to higher frequency due to the shrinkage of the ungated cavities in the lateral direction – the so-called side gating effect.

At small gate voltages, $|V_G| = 0 \ldots 0.5$ V, the concentration of the 2DEG is almost uniform across the whole grating period, and a delocalized resonant plasmon phase occurs. Here, the fundamental plasmon resonance for sample S13 is found at 1.72 THz. For sample S8, the third-order plasmon resonance occurs at 1.86 THz. Note that the plasmon resonances in the localized phase are stronger than in the delocalized phase.

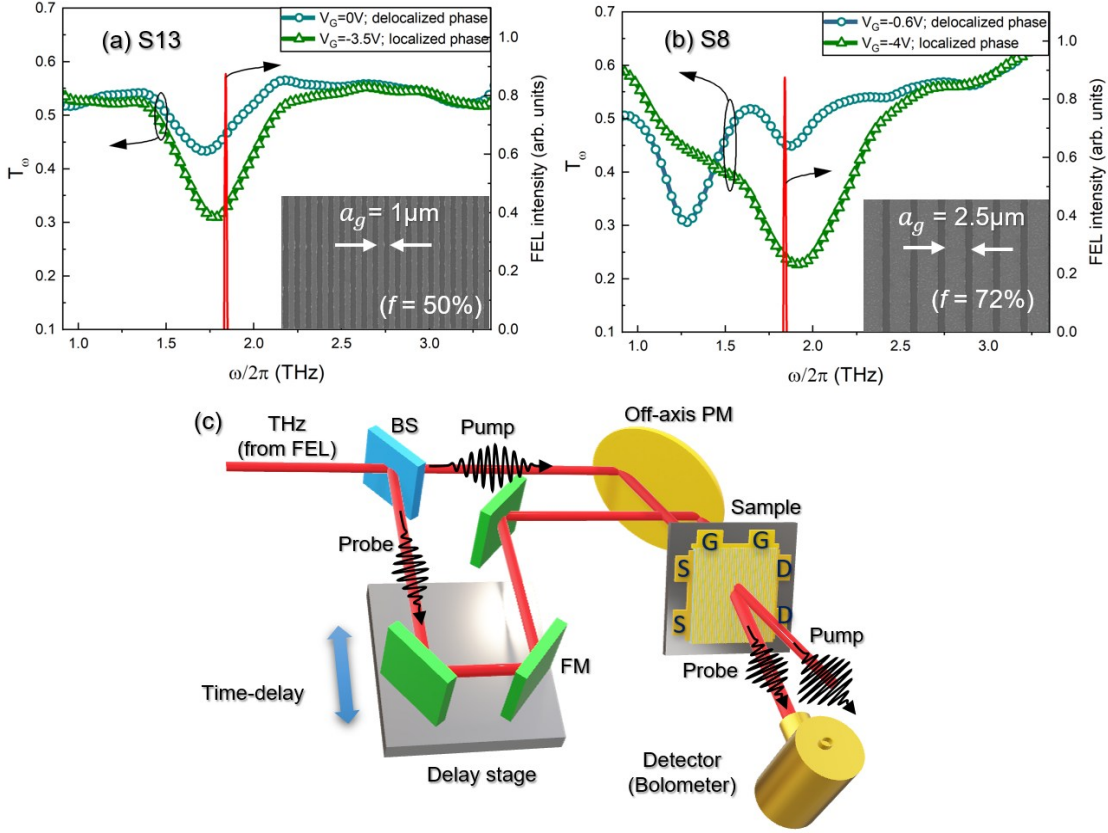

**Figure 1**. FTIR-measured transmittance spectra of samples S13 (a) and S8 (b) at different gate voltages. The red curves show the spectral line of the FEL. Scanning electron microscope images of grating fragments of the investigated samples are illustrated in the insets. The sketch of the THz pump-THz probe experiment (c), where BS is a beam splitter, PM is a parabolic mirror, and S, D, and G are the source, drain, and gate electrodes on the sample.

To study the nonlinear dynamics, pump-probe measurements were performed at the FEL facility – FELBE at the Helmholtz-Zentrum Dresden-Rossendorf.[35] The FEL was tuned to about 1.84 THz (located within the spectral band of all discussed plasmon resonances) with a spectral width of about 0.04 THz (see red line in Figure 1(a) and (b)) that is much narrower than plasmon resonant bands. The experimental setup is sketched in Figure 1(c). A small portion of the FEL radiation was split off to serve as a probe, and the main part of the FEL pulse energy was exploited to excite the sample. Both beams were focused on the same spot on the sample by an off-axis parabolic mirror to a spot diameter of about 0.8 mm. Pump and probe THz pulses were linearly co-polarized, and the polarization was set to be perpendicular to the finger structure of the grating, enabling excitation of plasmons. A delay stage in the probe beam path controlled the time delay, $\Delta t$, between pump and probe pulses. The integral transmission of the probe pulse, $T$, (the portion of the total energy of the probe pulse transmitted through the structure) was measured with a bolometer as a function of the time delay for various pump fluences, $F$.

In **Figure 2**, we show the pump-induced relative change in transmission, $\Delta T/T = [T(F) - T(0)]/T(F)$, for excitation fluences in the range of 1.5 nJ cm$^{-2}$ to 350 nJ cm$^{-2}$. As seen, the peak values of $\Delta T/T$ vary drastically when one compares the excitation of different plasmonic modes. However, the general form of this dependence is very similar, showing a decay time of the pump-probe signal of about 15 ps, which indicates similar cooling rates of the 2DEG in all cases. Panels (a) and (b) of Figure 2 refer to sample S13 and show the results for the excitation of the fundamental plasmon resonances in the delocalized and localized PC phases, respectively. In the former case, the relative transparency reaches a value of 5 % and is twice as large in the latter case. Panels (c) and (d) refer to the sample S8.

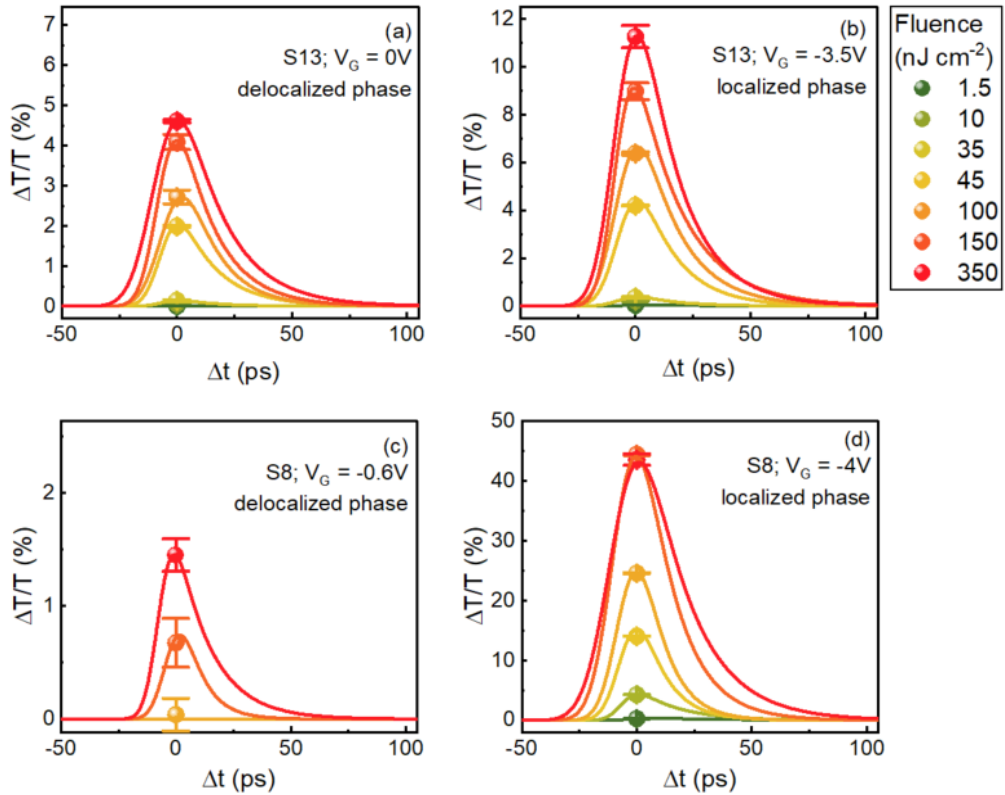

**Figure 2**. The relative pump-induced changes in transmission, $\Delta T/T$, measured at different pump pulse fluencies varied from 1.5 to 350 nJ cm$^{-2}$ (a)-(d). The corresponding samples and applied gate voltage are indicated in panels. Dots denote the peak values of $\Delta T/T$ taken at $\Delta t = 0$ ps.

The smallest response ($\Delta T/T \sim 1.5$ %) is observed for the third-order delocalized mode shown in Figure 2 (c), while the strongest response ($\Delta T/T \sim 45$ %) is observed for the resonant excitation of the fundamental localized mode shown in Figure 2 (d). The magnitudes of the pump-induced changes in the transmission correlate with the intensity of particular plasmon resonances observed in transmission spectra (see Figure 1 (a) and (b)), which were measured by FTIR spectroscopy in the linear regime. Indeed, the pump-induced transmission change becomes stronger when pumping the plasmon resonances of higher quality factors, that are observed in the localized phase of PC. Moreover, the near-field effects are stronger (see Ref. [33]) which leads to a stronger heating of the 2DEG than for the case of the delocalized plasmonic phase (see Section 3.2). All of these are the key factors of such a strong pump-induced transparency of the PC structures with an unprecedented value for a single channel of 2DEGs, even at such a low pump fluence far below 1 µJ cm$^{-2}$.

To further investigate the mechanisms behind the strong nonlinear absorption, we performed pump-probe experiments with a constant fluence, but varied the gate voltage, $V_G$. The results of these measurements are shown in **Figure 3** (a) and (b) for samples S13 and S8, respectively, by the color maps of relative transparency, $\Delta T/T$, as a function of both $\Delta t$ and $V_G$.

For an easier interpretation of these results, the corresponding color maps of linear transmittance as a function of the frequency and the gate voltage are presented in Figure 3 (c) and (d). Not surprisingly, the pump-probe signals are strongest when the carrier frequency of the FEL is situated within the excitation band of the particular plasmonic modes. This is realized when the vertical dashed line crosses the blue islands in Figure 3 (c) or (d). For example, for sample S13, the FEL frequency is in the plasmonic resonant bands at $-1\ V < V_G < 0.5\ V$ (excitation of the fundamental delocalized plasmon mode) and $-4\ V < V_G < -3\ V$ (excitation of the localized plasmon mode). In these ranges of $V_G$, a non-zero pump-probe signal, $\Delta T/T$ is identified (see red/blue islands in Figure 3 (a)). In the range of $-3\ V < V_G < -1\ V$, (out of the plasmonic bands) the non-linear response is absent, $\Delta T/T \sim 0$.

Figure 3 reveals the complex interplay of states with a positive and negative change in transmission, corresponding to pump-induced transparency and absorption, respectively. This unusual behavior can be explained by the co-existence of two effects, resulting from pump-induced strong electron heating. The first effect is a transient increase of the electron scattering rate that decreases the quality factor of the plasmon resonance. This leads to a weaker and broader plasmonic resonance. The second is the essential increase of the electron temperature, causing the redshift of the plasmon resonant bands. The latter can be associated with the increase of the effective electron mass.

While at resonance, the pump-probe signals are positive, which corresponds to a pump-induced transparency, we observe negative pump-probe signals outside the resonances. These negative signals mean a decrease in the transmission caused by the broadening and the red shift of the plasmonic absorption. For example, for sample S13 (Figure 3 (a)), the FEL excitation of the delocalized plasmon is accompanied by the emergence of a state with $\Delta T/T > 0$. The FEL excitation of a localized plasmon is accompanied by the emergence of a state with $\Delta T/T > 0$ at $-4\,V < V_G < -3.5\,V$ with gradual transition to the state with $\Delta T/T < 0$ at $V_G < -4\,V$. A similar evolution of the relative transparency associated with the localized plasmon resonance is observed for sample S8 (see panel (b)). However, the discussed non-linear effects of the transmittance modulation become stronger as a result of the initially sharper plasmon resonances (see panel(d)). On the contrary, for sample S8, the effects associated with the excitation of the delocalized plasmon of 3-rd order are weak and barely visible in panel (b).

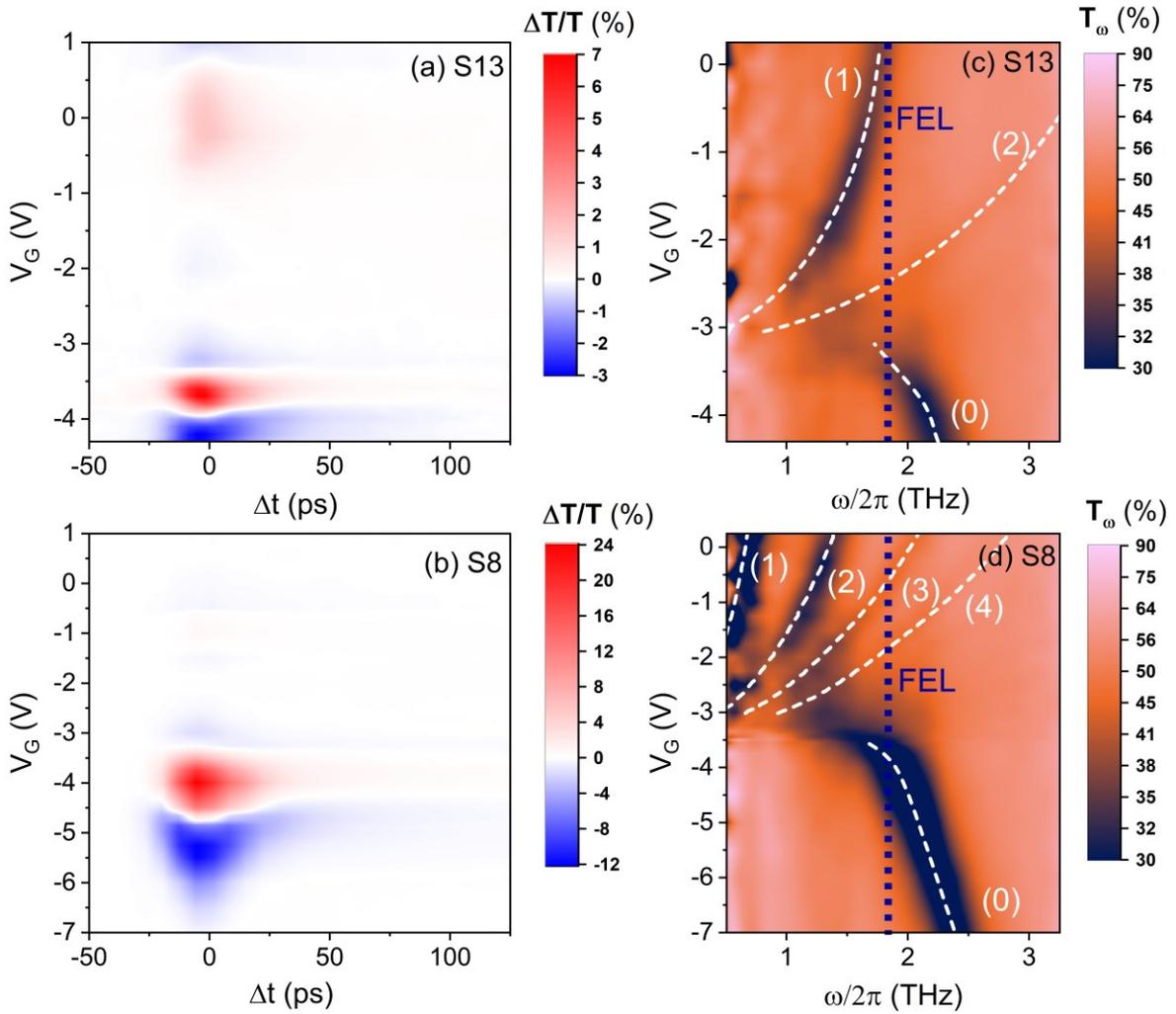

**Figure 3.** Contour plots of $\Delta T/T$ as a function of the gate voltage, $V_G$, and delay time, $\Delta t$, for samples S13 (a) and S8 (b). Contour plots of FTIR transmittance (measured in linear regime) as a function of the frequency and gate voltage for the sample S13 (c) and S8 (d). The blue dotted lines show the FEL carrier frequency. Dashed white curves show the positions of the transmission minima corresponding to the excitation of multiple-order delocalized (1-4) and localized (0) plasmon resonances.

Apparently, the asymmetry of the blue areas, i.e. the difference of the blue areas above and below resonance in Figure 3 (a) and (b), is a clear signature of the redshift of plasmon resonance rather than its broadening. It should be noted that this experiment also reveals the influence of the temperature on the effective electron mass: only the electron temperature is relevant and the equilibrium temperature seems to have only a minor impact. A similar behavior of pump-induced THz transmission has been observed in graphene plasmonic samples,[15] where it is related to the density of states, which increases linearly for graphene and stays constant for parabolic 2DEGs. In the case of GaN-based PCs, such a redshift of the plasmon resonance can be related to the temperature-dependent electron effective mass. The CR spectroscopy confirmed a significant increase of nearly 30% of the electron effective mass in the bare AlGaN/GaN sample in a wide temperature range of 10 - 300 K (see Section 1 of the Supporting Information).

In the final step, we investigated a nonlinear response of the fundamental delocalized plasmonic mode as a function of the equilibrium temperature, $T_0$. These results are shown in Figure 4. As can be seen, the pump-induced change in transmission stays nearly constant up to 40 K before it decreases linearly and nearly vanishes at a temperature of 160 K.

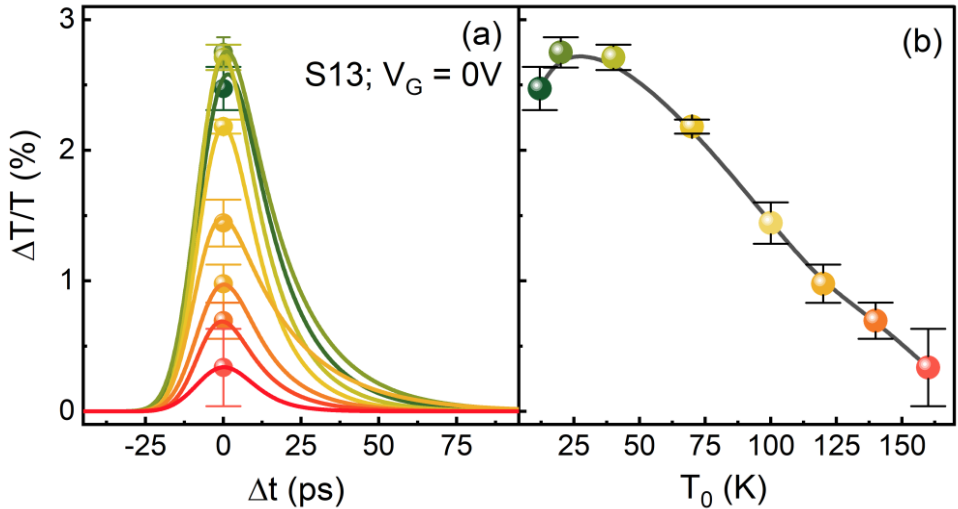

**Figure 4**. Measured time delay dependence of $\Delta T/T$ of the sample S13 in the delocalized plasmonic phase at different equilibrium temperatures, $T_0$. (a). The circles denote the peak values of the pump-probe signal. Temperature dependence of $\Delta T/T$ peak values are taken at $\Delta t = 0$ ps (b). The error bars indicate the standard deviation of experimental measurements.

Indeed, the simulation of the transport characteristics of the 2DEG (see details in Section 3.1) performed for $T_0 = 150$ K demonstrates a strong suppression of the heating effect in the wide range of applied electric fields. The electron temperature remains almost constant and the drift velocity demonstrates a much weaker non-linearity than at the deeper cryogenic temperature ($T_0 = 10$ K). In addition, at $T_0 = 150$ K the radiation frequency and the actual plasmon band can be off resonance due to the effective electron mass increase.

3. Theoretical framework and analysis

The theoretical description of the interaction effects of the high-field pulses with a PC structure is based on coupled solutions of Maxwell's equations and nonlinear electron transport equations. The geometry of the problem is sketched in **Figure 5**. The modeling structure consists of a periodic metallic grating with the thickness $d_g$, period $a_g$, and the width of grating bars, $w_g$. The grating is deposited on the AlGaN barrier layer with a thickness $d_b$, and dielectric permittivity $\epsilon_b$. The 2DEG is located between the thick SiC substrate (of thickness $d_s$, and permittivity $\epsilon_s$) and the barrier layer. In calculations, the 2DEG is treated as a delta-thin conductive layer with an initially equilibrium profile of the electron concentration, $n_0(x)$, that depends on the gate voltage and describes the modulation of carrier density. The metal bars of the grating are assumed to be gold with a bulk conductivity $\sigma_M$. The whole structure is placed in a vacuum.

The structure is uniformly illuminated at normal incidence by a Gaussian-type *em* pulse with the electric field polarized perpendicular to the gate fingers to allow for plasmonic excitation. In the selected geometry (see Figure 5), the incident pulse far from the PC structure has nonzero $E_x$ and $H_y$ components:

$$E_{i,x} = H_{i,y} = E_{max}\cos\left(2\pi\nu_c\left(\frac{z}{c} - t\right)\right)\exp\left(-4\log2\left[\frac{\frac{z}{c}-t}{\delta t}\right]^2\right), \tag{1}$$

where $c$ is the speed of light in vacuum, $E_{max}$ is the pulse amplitude, $\nu_c$ is the carrier frequency, and $\delta t$ is the transient half-width. All formulas in this article are presented in the centimeter-gram-second (CGS) system of units. However, when presenting the final results of calculations and measurements, we use the corresponding System International (SI) system units for clarity.

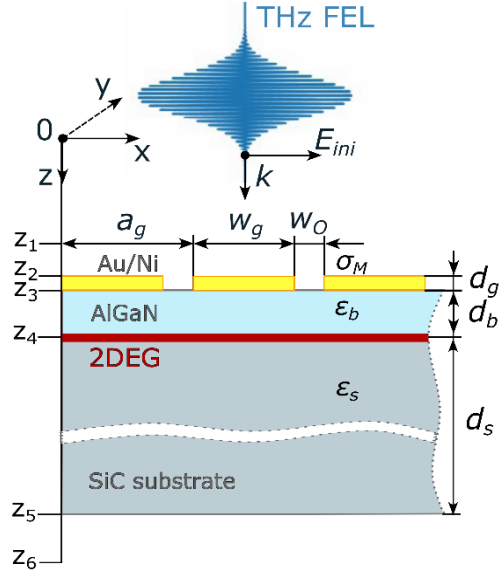

**Figure 5.** Model of the PC structure and geometry of the problem.

In the near-field zone of the PC structure ($z_2 - \Delta z_c < z < z_3 + \Delta z_c$, $\Delta z_c$ is of the order of a few periods), the resulting *em* pulse interacting with the plasmonic structure acquires a more complicated form, with the emergence of a non-zero z-component of the electric field. As a result, all actual components of the *em* field ($E_x, E_z, H_y$) become functions of both z and x coordinates, as well as time, $t$ and obey the following system:

$$\begin{cases} \frac{\epsilon(z)}{c}\frac{\partial E_x}{\partial t} = \frac{-\partial H_y}{\partial z} - \frac{4\pi}{c}[j_x^G(x,z,t) - j^{2D}(x,t)\delta(z-z_4)], \\ \frac{\epsilon(z)}{c}\frac{\partial E_z}{\partial t} = \frac{\partial H_y}{\partial x} - \frac{4\pi}{c}j_z^G(x,z,t), \\ \frac{1}{c}\frac{\partial H_y}{\partial t} = \frac{\partial E_z}{\partial x} - \frac{\partial E_x}{\partial z}. \end{cases} \quad (2)$$

Here, $j_{x,z}^G(x,z,t)$ are the $x$ and $z$ components of the conduction current density in the grating bars, and $j^{2D}(x,t)$ is the $x$ component of the 2D conduction a current-induced in the plane of 2DEG. We assume that dielectric permittivity, $\epsilon(z)$, has a step-like spatial profile, i.e., each layer has its dielectric permittivity. For the metallic grating, we assume Ohm's law, $j_{\{x,z\}}^G(x,z,t) = \sigma^G(x,z)E_{\{x,z\}}(x,z,t)$ with a step-like profile of the grating conductivity, $\sigma^G(x,z)$:

$$\sigma^G(x,z) = \begin{cases} \sigma_M \text{ at } z_2 < z < z_3, 0 < x < w_G \\ 0, \quad \text{otherwise} \end{cases}$$

Beyond the small-signal approach, the current, $j^{2D}(x,t)$, is no longer a linear function of the $x$ component of the electric field in the 2DEG plane, $E_x(x,z_4,t) \equiv E^{2D}(x,t)$.

To describe the spatio-temporal 2D electron dynamics, we used the so-called viscous HD model [46, 47] written for a 2DEG with Fermi statistics and parabolic electron dispersion law described by the effective mass, $m^*$. The applicability of the used HD transport model is discussed in Section 2 of the Supporting Information. The basic system of equations in the proposed transport model contains the continuity equation:

$$\frac{\partial n}{\partial t} + \frac{1}{m^*}\frac{\partial (nP_D)}{\partial x} = 0, \qquad (3)$$

the Navier-Stokes equation:

$$\frac{\partial P_D}{\partial t} + \frac{P_D}{m^*}\frac{\partial P_D}{\partial x} + \frac{C_V}{2m^*}\frac{\partial P_T^2}{\partial x} + \frac{\pi\hbar^2}{m^*(1-exp(\xi_0))}\frac{\partial n}{\partial x} - \frac{e\partial\varphi_0}{\partial x} - \frac{\nu}{m^*}\frac{\partial^2 P_D}{\partial x^2} = -eE^{2D}(x,t) + Q_p(x,t), \quad (4)$$

and the heat transport equation:

$$\frac{\partial P_T^2}{\partial t} + \frac{1}{m^*}\frac{\partial P_D P_T^2}{\partial x} - \frac{\chi_V}{C_V}\frac{\partial^2 P_T^2}{\partial x^2} - \frac{\nu}{C_V m^*}\left(\frac{\partial P_D}{\partial x}\right)^2 = \frac{(Q_E(x,t)-2P_D Q_P(x,t))}{C_V}. \qquad (5)$$

The system (3)-(5) is formulated for the 2DEG concentration, $n(x,t)$, the electron drift momentum $P_D(x,t) = m^*V_D(x,t)$, that corresponds to the electron drift velocity, $V_D$, and the thermal electron momentum, $P_T(x,t) = \sqrt{2m^*k_B T_e(x,t)}$, where $T_e$ is the electron temperature. The $\varphi_0$ is the built-in electrostatic potential which defines the equilibrium lateral profile, $n_0(x)$, of the 2DEG concentration, $\varphi_0 = \frac{k_B T_0}{e}\ln[\exp(n_0(x)/N_c) - 1]$, $N_c = m^*k_B T_0/\pi\hbar^2$, where $k_B$ and $e$ are the Boltzmann constant and elementary charge, respectively, $T_0$ is the equilibrium temperature. The factor $C_V$ denotes the dimensionless heat capacity of the 2DEG. For the degenerate case, $C_V = \frac{2F_1(\xi_0)}{\xi_0} - \frac{\xi_0}{1-exp(\xi_0)}$, where $\xi_0 = 2\pi\hbar^2 n_0/P_T^2$ is the function of the dimensionless Fermi level and $F_1(\xi_0) = \int_0^\infty \frac{xdx}{1+exp(x-\xi_0)}$ is the Fermi integral. The factors $\nu$ and $\chi_V$ are the coefficients of the kinematic viscosity and normalized thermal conductivity of the 2DEG, respectively. In the low temperature limit, $T_e \ll T_F$, (where $T_F = \frac{E_F}{k_B} = \pi\hbar^2 n/m^* k_B$ is the Fermi temperature) these factors can be estimated as follows (see Ref. [36]:

$$\nu \approx \frac{2\hbar}{\pi m^*}(T_F/T_e)^2/\ln(2T_F/T_e) \text{ and } \chi_V \approx \frac{4\pi\hbar}{3m^*}(T_F/T_e)\ln(2T_F/T_e) \qquad (6)$$

The frames of the applicability of the viscous hydrodynamic model are briefly discussed in the Supporting Information (see Section 2).

Finally, functions $Q_p$ and $Q_E$ are the momentum and energy dissipation rates. In our model, $Q_{p,E} = Q_{p,E}^{qel} + Q_{p,E}^{pop}$ contains two terms responsible for the relaxation of the electron subsystem due to quasi-elastic scattering mechanisms and a non-elastic mechanism of the electron-longitudinal optical (LO) phonon scattering. The former terms are taken in the simplified form (without specification of particular scattering mechanisms): $Q_p^{qel} = -P_D/\tau_p$, $Q_E^{qel} = -(P_T^2 - P_{T_0}^2)/\tau_E$, where $\tau_p$ and $\tau_E$ are the momentum and energy relaxation times corresponding to quasi-elastic mechanisms and $P_{T_0}^2 = \sqrt{2m^*k_B T_0}$. The contributions coming from the electron-LO scattering become important and they can lead to the nonlinearity of transport characteristics. In this paper, we used the explicit forms of terms $Q_{p,E}^{pop}$ given in Ref. [48].

Finally, Maxwell's equations (2) and the HD transport equations (3)-(5) are coupled through the definition of the current density,

$$j^{2D}(x,t) = \frac{-en(x,t)P_D(x,t)}{m^*}. \tag{7}$$

Thus, the theoretical approach is reduced to solving a system of nonlinear partial differential equations (2)-(5) and (7). This system was solved by the finite-difference time-domain (FDTD) simulation scheme using the so-called Leapfrog algorithm [49] with a proper treatment of the delta-thin conductive layer [43, 44] and a particular discretization of the HD transport Equations (3)-(5). Also, the periodical boundary conditions with respect to the grating period $a_G$ are assumed for all input variables along x-direction.

The FDTD simulation allows us to calculate the optical characteristics of the considered PC structures. Particularly, having the transient waveform of the electric component, $E_{ref}(t) \equiv E_{i,x}(z_1, t)$ of the incident pulse (recorded at some point far above the PC structure, say at the coordinate $z_1$), we can find a transient waveform of the actual electric component of the transmitted pulse (recorded at some point below the PC structure, say at the coordinate $z_5$), $E_{tran}(t) \equiv E_x(z_5, t)$. Using the Fourier transform of both pulses in the particular transient window, $[t_0, t_1]$, we can find the complex-valued spectral component of both pulses, $E_{ref}(\omega)$ and $E_{tran}(\omega)$, and calculate the spectral transmission coefficient, $T_\omega$:

$$T_\omega = |E_{tran}(\omega)/E_{ref}(\omega)|^2. \tag{8}$$

We will also analyze the integral transmission, $T$, defined as follows:

$$T = \int_{t_0}^{t_1} dt\, [E_{tran}(t)]^2 / \int_{t_0}^{t_1} dt [E_{ref}(t)]^2. \tag{9}$$

It determines the portion of the incident pulse energy transmitted through the structure. In the processing of the experimental results, we will use relative transparency, $\Delta T/T(E_{max})$, as a function of the incident pulse amplitude (or the energetic fluence). This quantity will be defined as follows:

$$\Delta T/T(E_{max}) = \frac{[T(E_{max}) - T(E_{0,max})]}{T(E_{max})}. \tag{10}$$

Here, $E_{0,max}$ is a small amplitude of the incident pulse when the linear response of the PC structure occurs.

The results of the FDTD simulation of relative transparency are presented in Section 3.2. We note that rigorous pump-probe spatio-temporal simulations involving two pulses of different amplitudes with an oblique incidence of one of them is a much more difficult problem that is beyond the scope of this work. The calculated characteristics $\Delta T/T$, concerning only pump pulse, describe the results of the pump-probe measurements well and highlight the main physical effects arising at high-field resonant excitations of plasmons in a 2DEG.

3.1. Steady-state transport characteristics

To illustrate the expected temperature range of the electron heating due to the interaction with pump beam, we performed a simulation of the steady-state (DC) transport characteristics in a uniform electric field (see **Figure 6**). At low temperatures and small DC electric fields, $E_{dc} < 0.5$ kV cm$^{-1}$, the drift velocity (curve 1$^a$) is still a linear function of the electric field with the slope defined by low field mobility, $\mu_0 = 7500$ cm$^2$ V$^{-1}$ s$^{-1}$ ($\tau_p = 0.95$ ps at $m^* = 0.22 \times m_e$ and $\tau_E = 10$ ps). Meanwhile, the electron temperature (curve 1$^b$) rapidly increases to ~100 K. Starting from 1 kV cm$^{-1}$, the electron-LO phonon scattering is activated. It leads to a sub-linear behavior of the field dependence $V_D(E_{dc})$, and stabilization of the electron temperature growth in the 120-150 K range.

For high temperatures, $T_0 = 150$ K, the main scattering mechanism is the intensive electron-LO phonon scattering. As a result, the 2DEG remains near the equilibrium in a wide range of the applied field, with almost constant electron temperature, $T_e = T_0$ (curve 2$^b$) and a much weaker dependence of $V_D(E_{dc})$ (curve 2$^a$).

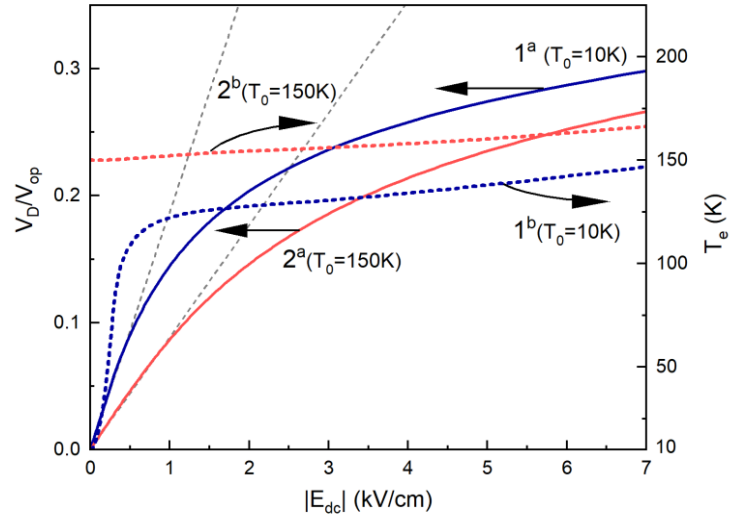

**Figure 6.** Steady-state field dependences of the drift velocity (curves 1$^a$ and 2$^a$) and the electron temperature (1$^b$ and 2$^b$) for two equilibrium temperatures $T_0 = 10$ K and 150 K, respectively. The characteristic velocity $V_{op} = \sqrt{2\,\epsilon_{LO}/m^*} = 3.83 \times 10^7$ cm s$^{-1}$, where $\epsilon_{LO}$ is the LO-phonon energy. Calculations are performed without taking into account the effective electron mass increase.

It should be noted that the magnitude of the electron gas heating in a DC applied electric field and under the action of an AC electromagnetic pulse should be different, and less in the latter case due to the alternating behavior of the electric component in the pulse. However, we expect that pump pulses with amplitudes, $E_{max}$, well larger than 1 kV cm$^{-1}$ can already induce transient changes in transport characteristics of the electron gas and, thereby, self-consistently modify the optical characteristics of the PC structure. The latter strongly depends on the pulse parameters (duration and carrier frequency), and initial parameters of the 2DEG including the geometry of the PC structure. Moreover, spatial non-uniformity of the electric field, $E^{2D}(x,t)$, acting on the 2DEG induces an additional effect on transmission spectra coming from gradient terms provided by the viscous HD transport model (see Eqs. (3)-(5)).

## 3.2. Analysis of nonlinear effects

After confirming that the developed theoretical model and FDTD algorithm accurately reproduce the results of FTIR THz characterization in the linear regime (see Section 3 in the Supporting Information), they were applied to simulate the nonlinear interaction of a relatively long, large-amplitude *em* pulse with the PC structure. We consider sample S13 and the incident pulse with the following parameters: $\delta t = 16\ ps$ and $E_{\max} = 3$ kV cm$^{-1}$, which is provided by FELBE. The result of calculations of the transient waveform of the incident and transmitted pulses and their power Fourier spectra are shown in **Figure 7** (a) and (b), respectively. Also, snapshots of the spatial form of the *em* pulse interacting with the PC structure are shown at different times. For this case, the spatial half-width of the pulse is approximately 0.5 cm (see Figure 7 (c)), which is much larger than the substrate thickness. As a result, the effect of echoes vanishes in the transmitted pulse (see red curve in Figure 7 (a)).

Note that this simulation requires a wider transient window [0, 70 ps] (t$_1$ = 70 ps) than in the linear regime (see section 3 in the Supporting Information). As can be seen, the transmitted pulse becomes wider in space or much narrower in the frequency domain (see the red curve in Figure 7 (b)). The transmitted field amplitude is half that of the incident one. This is due to the reflection of the pulse from the structure (the formation of the reflected pulse is seen in panels Figure 7 (e)-(g)) and a partial absorption of *em* energy by the 2D plasmons. It should be noted that due to the 2D plasmon resonance, the Fourier spectral characteristic of the transmitted pulse (see the red curve in Figure 7 (b)) experiences a deviation from the purely Gaussian shape of the incident pulse. Also, we observe a visible modification of the spectral components of the large-amplitude transmitted pulse calculated at $E_{\max} = 3$ kV cm$^{-1}$.

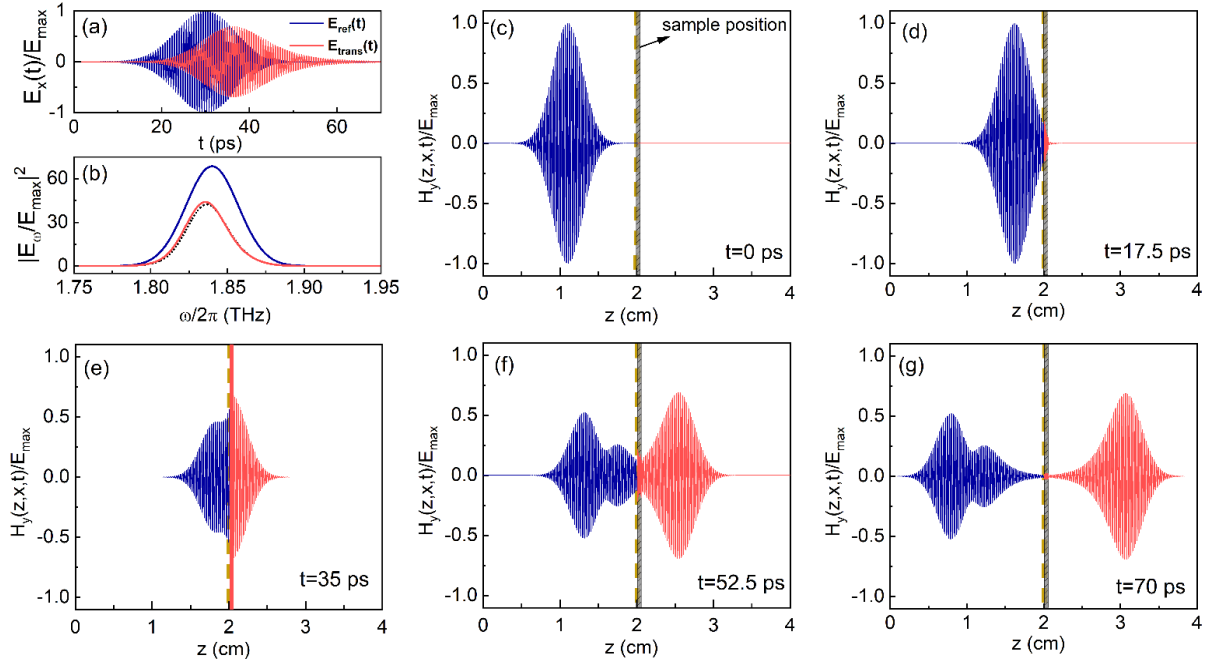

**Figure 7.** The shape of the incident and transmitted pulses in the time domain calculated at $E_{max} = 3$ kV cm$^{-1}$, $\delta t = 16$ ps, and $\nu_c = 1.84$ THz (a). The corresponding normalized power Fourier spectrum (dashed black line) is calculated at a small field, $E_{max} = 0.1$ kV cm$^{-1}$ (b). The snapshots of the spatial distribution of the magnetic components of the FEL pulse (c)-(g), that interact with PC structure at different moments. The incoming and reflected signals are shown with blue curves, and red curves show the transmitted signal. The PC sample is illustrated in the background of the data in (c)-(g) as a vertical bar at $z = 2$ cm, where the sample thickness is shown in the scale, but the grating electrodes are not.

The effects of electron heating under the action of the large amplitude pulse are illustrated in **Figure 8**. The electric fields acting on the 2DEG can reach values of ~6 kV cm$^{-1}$, twice larger than the incident pulse peak electric field. This is a result of the near-field effects induced by the metallic gratings (see Ref. [33, 42]), a spatial pulse compression due to the refractive index of the substrate, and resonant excitations of the 2D plasmons. The electrons can be accelerated to drift velocities of 0.1 - 0.15 $V_{op}$(~0.5×10$^7$ cm s$^{-1}$) together with heating up to an electron temperature of 100 - 120 K. Note that the steady-state calculations (see curves 1$^a$ and 1$^b$ in Figure 6) predict the onset of a nonlinear relationship between the current induced in the 2DEG and electric fields. Also, the CR measurements show a 7 % increase in the electron effective mass with respect to the equilibrium value at 10 K (see Section 1 in the Supporting Information). All of these lead to an increased transparency of the PC structure. Particularly, the integral transmission $T$ calculated at $E_{max} = 3$ kV cm$^{-1}$ is equal to 0.53 and the low-field integral transmission $T$ calculated at $E_{max} = 0.1$ kV cm$^{-1}$ is equal to 0.505. This gives the relative transparency $\Delta T/T = 4.7$ %, while the pump-probe measurements give a slightly smaller value of relative transparency ~ 4 % (see **Figure 9** (a)).

The nonlinear effect of a self-induced transparency becomes larger in the case of the localized PC phase. In this phase, only the ungated part of the 2DEG is the active element of the whole plasmonic structure which effectively absorbs *em* radiation due to a resonant interaction with the localized 2D plasmon excited in the electrically-formed 2DEG-strip grating (see Ref. [33, 38]). Here, the effects of the electron gas heating become larger due to the stronger near-field effects. This is illustrated in Figure 8 (d-f) as an example of spatial distributions of the electric field in the plane of 2DEG and electron transport characteristics.

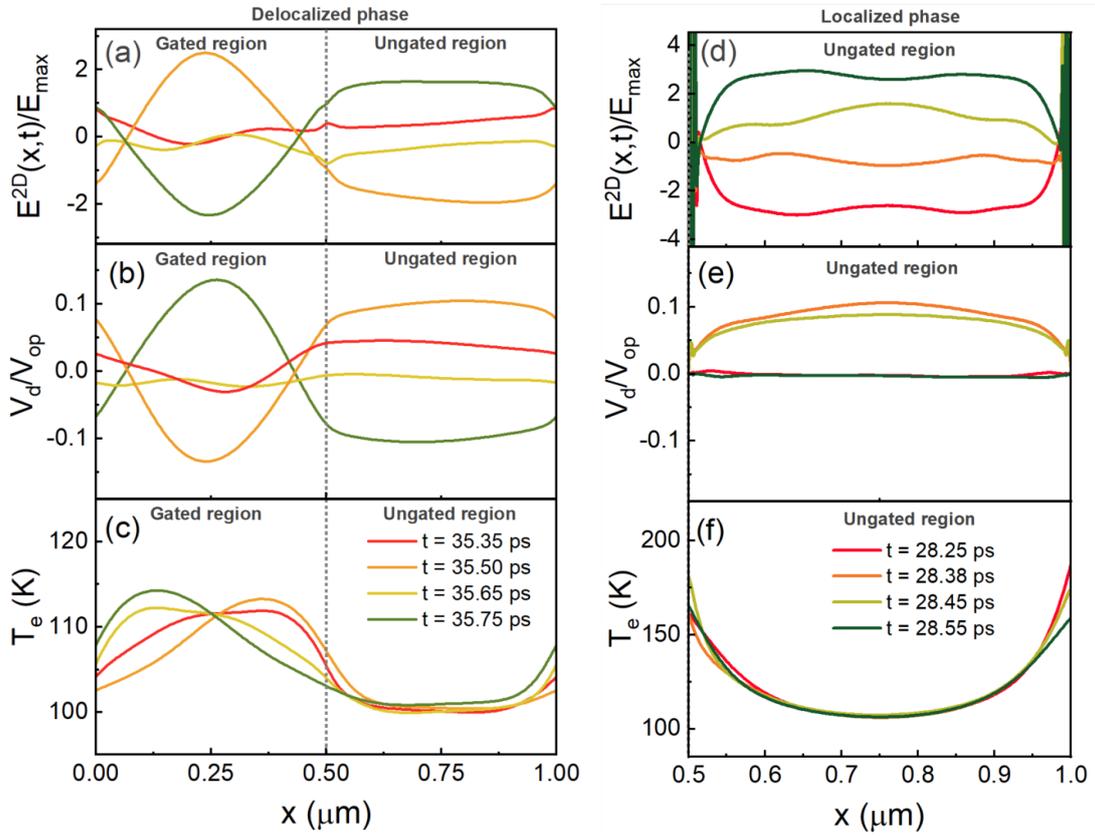

**Figure 8.** A spatial distribution of the electric field in the 2DEG plane of one period of the PC (a) and (d), the electron drift velocity (b) and (e), and the electron temperature (c) and (f) of the 2DEG at four different times. Results are shown for the delocalized phase, $V_G = 0$ V, (a-c) and the localized phase at $V_G = -3.5$ V and $E_{max} = 2$ kV cm$^{-1}$ (d-f).

As shown in Figure 8 (d), the amplitudes of the electric fields acting on the 2DEG can be up to 3 times larger than the peak field of the incident pulse. As a result, the similar heating of the 2DEG in the localized phase can be reached at a smaller $E_{max}$ than in the delocalized phase.

Moreover, at the edges of the ungated region, $T_e$ reaches a value of 150-180 K. This can lead to a stronger increase of the effective mass and thus a redshift of the plasmon frequency. Calculations give $T = 0.318$ at $E_{max} = 2$ kV cm$^{-1}$ and $T = 0.295$ at a small field, $E_{max} = 0.1$ kV cm$^{-1}$. This gives a relative increase in transparency of 7%, well-correlated with measured values of ~ 5-6%.

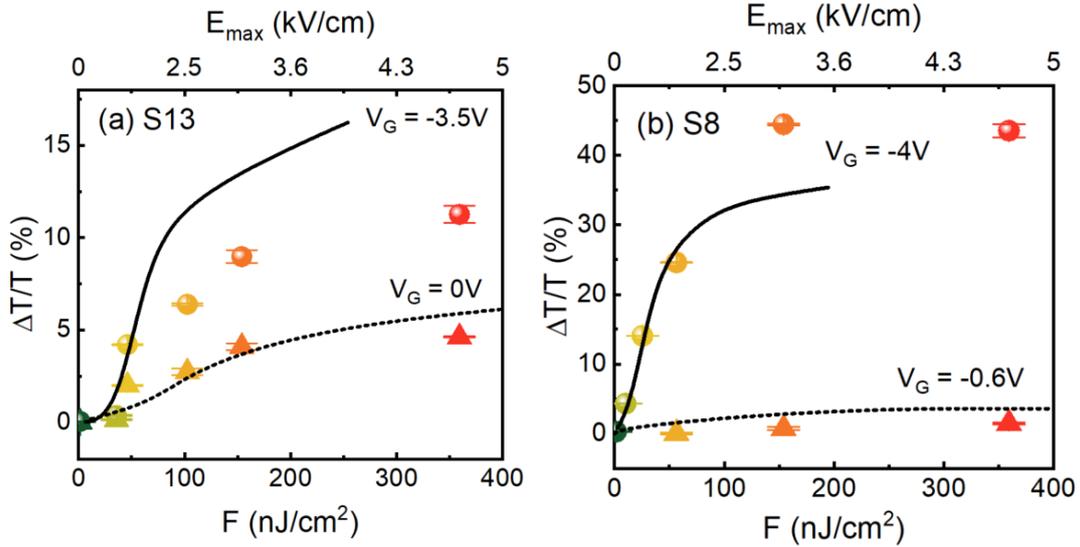

**Figure 9**. The calculated fluence dependences of $\Delta T/T$ characteristic at $\Delta t = 0$ ps for samples S13 (a) and S8 (b). Data points represent experimental values taken from the peaks in Figure 2).

Additionally, we performed FDTD simulations of the optical characteristics of another sample - S8. The nonlinear effect of a self-induced relative transparency in the localized phase for this sample is 3 times weaker than for S13 in the same phase and does not exceed 2%. This is due to the excitation of an initially weak plasmon resonance of the 3-rd order. However, in the localized phase the discussed effect is the strongest among all cases. For example, in the calculations, the integral transmission, $T = 0.226$ in the linear regime ($E_{max} = 0.01$ kV cm$^{-1}$), and 0.344 in the nonlinear regime ($E_{max} = 3$ kV cm$^{-1}$). This gives the unprecedented value of a relative transparency ~ 35% (see Figure 9 (b)).

All simulated curves in Figure 9 show a specific behavior with the saturation that can be explained by the saturation of the electron temperature and a sublinear behavior of the drift velocity at electric fields well above 1 kV cm$^{-1}$, as shown in Figure 6. Remarkably, the results of a simulation performed using the proposed transport model agree well with the results of the measurements. It should be noted that the effects of the transparency changes disappear with an increase of the equilibrium temperature, $T_0$ (see Figure 4). This is the result of a general decrease of the plasmon resonances quality factor and almost linear response (a constant electron temperature and an Ohmic dependence of $V_d(E)$ ) of the 2DEG in the considered range of electric fields (see curves $2^a$ and $2^b$ in Figure 6).

## 4. Discussion and conclusions

THz nonlinear materials are essential for a variety of applications such as saturable absorbers or harmonic generation, though the options for choosing the appropriate system are rather limited.[50] Certain bulk semiconductors with a well-matched doping level have shown a pump-induced increase of transmission of the order of 10% when excited with a field strength in the 100 kV cm$^{-1}$ range.[51] While this nonlinearity might be suitable for specific applications, it offers minimal flexibility or tunability. One route to achieve a tunable nonlinearity is the exploitation of a 2DEG, in which the electron density can be tuned by the field effect. This approach usually leads to a weak interaction between THz radiation and the electrons in the 2DEG due to the small interaction volume, which can be improved by guiding the THz radiation along the 2DEG.[52] Graphene has been proven to offer an exceptionally strong THz nonlinear absorption resulting in a saturation of the absorption in the 10% range under normal incidence when illuminated with THz pulses with a field strength of several tens of kV cm$^{-1}$. As has been shown, this nonlinearity results in a record-strength harmonic generation.[6] Plasmonic structures offer a straightforward way to increase the light-matter interaction by patterning the materials: bulk InAs patterned into plasmonic discs have shown a rather strong nonlinear response, leading to a change in transmission of the order of 50%, though tuning of the concentration is not possible in such a system.[53] In recent years, the combination of 2DEG- and graphene-based structures with plasmon resonances has been demonstrated in the 10% range of THz transmission modulation.[15, 54]

As opposed to the graphene plasmonic studies, the situation in our case is very different: while graphene features a linear and gapless band structure and a linearly increasing density of states, the 2DEG in our heterostructure is characterized by parabolic bands and a direct band gap. For the graphene plasmonics, the linearly increasing density of states leads to a decrease of the chemical potential with increased electron temperature. The thermal effect dominates the nonlinear THz response in the case of graphene. For our samples, hosting a 2DEG with parabolic dispersion, with a constant density of states, an increase in the electron temperature would not lead to a change in chemical potential. But as the effective mass, and thus the density of states, changes significantly with increasing temperature, we observe the same type of redshift as for the graphene case.

The pump-induced increase in transmission observed in our study offers an unprecedentedly strong THz nonlinear transmission for the 2DEG of up to 45% at a fluence of less than 400 nJ cm$^{-2}$, which corresponds to a field strength of about 5 kV cm$^{-1}$. At the same time, the gate electric field effect enables tunability of plasmons in a wide frequency range. By tuning the 2DEG concentration, the grating gate allows to switch from pump-induced transmission to no nonlinearity at all and further to pump-induced absorption. Additionally, the metallic grating generates strong near-field effects concentrating the electric field near the active 2DEG layer, which further amplifies THz nonlinearity, particularly in the case of the localized PC phase, where the near-field effects are strong. Furthermore, the AlGaN/GaN-based heterostructures are well known for their durability even at high power density, which makes the demonstrated PC a very flexible device for THz nonlinear applications such as saturable absorbers.

We have developed a theoretical model that can be applied to the analysis of the effects of a nonlinear interaction of high-field and ps duration pulses with the 2DEG under the metallic gratings. The approach was based on FDTD solutions (using the Leapfrog algorithm) of a system of Maxwell's equations (written for normal incidence and the electric field polarized perpendicular to the gate fingers) coupled with the viscous HD transport model of the electron kinetics. The approach takes into account the effects of the non-uniform electron heating of the 2DEG under the grating, nonlinear field dependences of basic transport characteristics such as the drift velocity and the electron temperature including the electron temperature dependencies of viscous parameters of the 2DEG and the electron effective mass. The proposed theoretical model reproduces well the results of THz pump-THz probe measurements and provides quantitative explanations of the observed nonlinear phenomena. In general, the developed simulation algorithm can be applied for the investigation of both linear and nonlinear THz responses of various PC structures with different grating geometry and different basic parameters of the 2DEG formed in polar semiconductor heterostructures.

**Supporting Information**

Supporting Information is available from the Wiley Online Library or from the author.

Acknowledgements

The work was supported by the European Union through ERC-ADVANCED grant TERAPLASM (No. 101053716). Views and opinions expressed are, however, those of the authors only and do not necessarily reflect those of the European Union or the European Research Council Executive Agency. Neither the European Union nor the granting authority can be held responsible for them.21

We also acknowledge the support of "Center for Terahertz Research and Applications (CENTERA2)" project (FENG.02.01-IP.05-T004/23) carried out within the "International Research Agendas" program of the Foundation for Polish Science co-financed by the European Union under European Funds for a Smart Economy Programme. The research leading to this result has been co-funded by the project NEPHEWS under Grant Agreement No 101131414 from the EU Framework Programme for Research and Innovation Horizon Europe. V. K., and S. K. acknowledge the long-term program of support of the Ukrainian research teams at the Polish Academy of Sciences carried out in collaboration with the U.S. National Academy of Sciences with the financial support of external partners. M.M. acknowledges funding by the Deutsche Forschungsgemeinschaft (DFG, German Research Foundation)—Project-ID 278162697—SFB1242. We thank J. Michael Klopf and the ELBE team for their assistance.

Received: ((will be filled in by the editorial staff))
Revised: ((will be filled in by the editorial staff))
Published online: ((will be filled in by the editorial staff))

References
[1] V. Berger, Nonlinear Photonic Crystals, Physical Review Letters, 81 (1998) 4136-4139.
[2] N. Yoshikawa, T. Tamaya, K. Tanaka, High-harmonic generation in graphene enhanced by elliptically polarized light excitation, Science, 356 (2017) 736-738.
[3] J. Raab, F.P. Mezzapesa, L. Viti, N. Dessmann, L.K. Diebel, L. Li, A.G. Davies, E.H. Linfield, C. Lange, R. Huber, M.S. Vitiello, Ultrafast terahertz saturable absorbers using tailored intersubband polaritons, Nature Communications, 11 (2020) 4290.
[4] M.R. Shcherbakov, P.P. Vabishchevich, A.S. Shorokhov, K.E. Chong, D.-Y. Choi, I. Staude, A.E. Miroshnichenko, D.N. Neshev, A.A. Fedyanin, Y.S. Kivshar, Ultrafast All-Optical Switching with Magnetic Resonances in Nonlinear Dielectric Nanostructures, Nano Letters, 15 (2015) 6985-6990.
[5] Z. Chai, X. Hu, F. Wang, X. Niu, J. Xie, Q. Gong, Ultrafast All-Optical Switching, Advanced Optical Materials, 5 (2017) 1600665.
[6] H.A. Hafez, S. Kovalev, J.-C. Deinert, Z. Mics, B. Green, N. Awari, M. Chen, S. Germanskiy, U. Lehnert, J. Teichert, Extremely efficient terahertz high-harmonic generation in graphene by hot Dirac fermions, Nature, 561 (2018) 507-511.

[7] Z. Mics, K.-J. Tielrooij, K. Parvez, S.A. Jensen, I. Ivanov, X. Feng, K. Müllen, M. Bonn, D. Turchinovich, Thermodynamic picture of ultrafast charge transport in graphene, Nature Communications, 6 (2015) 7655.

[8] S. Kovalev, H.A. Hafez, K.-J. Tielrooij, J.-C. Deinert, I. Ilyakov, N. Awari, D. Alcaraz, K. Soundarapandian, D. Saleta, S. Germanskiy, and M. Chen, Electrical tunability of terahertz nonlinearity in graphene, Science advances, 7 (2021) eabf9809.

[9] J.C. Deinert, D. Alcaraz Iranzo, R. Pérez, X. Jia, H.A. Hafez, I. Ilyakov, N. Awari, M. Chen, M. Bawatna, A.N. Ponomaryov, S. Germanskiy, M. Bonn, F.H.L. Koppens, D. Turchinovich, M. Gensch, S. Kovalev, K.J. Tielrooij, Grating-Graphene Metamaterial as a Platform for Terahertz Nonlinear Photonics, ACS Nano, 15 (2021) 1145-1154.

[10] A. Maleki, M.B. Heindl, Y. Xin, R.W. Boyd, G. Herink, J.-M. Ménard, Strategies to enhance THz harmonic generation combining multilayered, gated, and metamaterial-based architectures, Light: Science & Applications, 14 (2025) 44.

[11] S.H. Lee, M. Choi, T.-T. Kim, S. Lee, M. Liu, X. Yin, H.K. Choi, S.S. Lee, C.-G. Choi, S.-Y. Choi, Switching terahertz waves with gate-controlled active graphene metamaterials, Nature materials, 11 (2012) 936-941.

[12] L. Wang, Y. Zhang, X. Guo, T. Chen, H. Liang, X. Hao, X. Hou, W. Kou, Y. Zhao, T. Zhou, A review of THz modulators with dynamic tunable metasurfaces, Nanomaterials, 9 (2019) 965.

[13] M. Kauranen, A.V. Zayats, Nonlinear plasmonics, Nature Photonics, 6 (2012) 737-748.

[14] M.M. Jadidi, J.C. König-Otto, S. Winnerl, A.B. Sushkov, H.D. Drew, T.E. Murphy, M. Mittendorff, Nonlinear terahertz absorption of graphene plasmons, Nano Letters, 16 (2016) 2734-2738.

[15] M.M. Jadidi, K.M. Daniels, R.L. Myers-Ward, D.K. Gaskill, J.C. König-Otto, S. Winnerl, A.B. Sushkov, H.D. Drew, T.E. Murphy, M. Mittendorff, Optical control of plasmonic hot carriers in graphene, ACS Photonics, 6 (2019) 302-307.

[16] M.L. Chin, S. Matschy, F. Stawitzki, J. Poojali, H.A. Hafez, D. Turchinovich, S. Winnerl, G. Kumar, R.L. Myers-Ward, M.T. Dejarld, Observation of strong magneto plasmonic nonlinearity in bilayer graphene discs, Journal of Physics: Photonics, 3 (2021) 01LT01.

[17] J.W. Han, M.L. Chin, S. Matschy, J. Poojali, A. Seidl, S. Winnerl, H.A. Hafez, D. Turchinovich, G. Kumar, R.L. Myers-Ward, Plasmonic Terahertz Nonlinearity in Graphene Disks, Advanced Photonics Research, 3 (2022) 2100218.

[18] G.C. Dyer, G.R. Aizin, S.J. Allen, A.D. Grine, D. Bethke, J.L. Reno, E.A. Shaner, Induced transparency by coupling of Tamm and defect states in tunable terahertz plasmonic crystals, Nature Photonics, 7 (2013) 925-930.

[19] A. Muravjov, D. Veksler, V. Popov, O. Polischuk, N. Pala, X. Hu, R. Gaska, H. Saxena, R. Peale, M. Shur, Temperature dependence of plasmonic terahertz absorption in grating-gate gallium-nitride transistor structures, Applied Physics Letters, 96 (2010) 042105.

[20] N. Nader Esfahani, R. Peale, W. Buchwald, C. Fredricksen, J. Hendrickson, J. Cleary, Millimeter-wave photoresponse due to excitation of two-dimensional plasmons in InGaAs/InP high-electron-mobility transistors, Journal of Applied Physics, 114 (2013).

[21] M. Białek, M. Czapkiewicz, J. Wróbel, V. Umansky, J. Łusakowski, Plasmon dispersions in high electron mobility terahertz detectors, Applied Physics Letters, 104 (2014).

[22] B. Yan, J. Fang, S. Qin, Y. Liu, Y. Zhou, R. Li, X.-A. Zhang, Experimental study of plasmon in a grating coupled graphene device with a resonant cavity, Applied Physics Letters, 107 (2015).

[23] S. Boubanga-Tombet, W. Knap, D. Yadav, A. Satou, D.B. But, V.V. Popov, I.V. Gorbenko, V. Kachorovskii, T. Otsuji, Room-Temperature Amplification of Terahertz Radiation by Grating-Gate Graphene Structures, Physical Review X, 10 (2020) 031004.

[24] Z. Fei, A. Rodin, G.O. Andreev, W. Bao, A. McLeod, M. Wagner, L. Zhang, Z. Zhao, M. Thiemens, G. Dominguez, Gate-tuning of graphene plasmons revealed by infrared nano-imaging, Nature, 487 (2012) 82-85.

[25] H. Qin, Y. Yu, X. Li, J. Sun, Y. Huang, Excitation of terahertz plasmon in two-dimensional electron gas, Terahertz Science and Technology, 9 (2016) 71.

[26] P. Alonso-González, A.Y. Nikitin, Y. Gao, A. Woessner, M.B. Lundeberg, A. Principi, N. Forcellini, W. Yan, S. Vélez, A.J. Huber, Acoustic terahertz graphene plasmons revealed by photocurrent nanoscopy, Nature nanotechnology, 12 (2017) 31-35.

[27] D. Pashnev, V.V. Korotyeyev, J. Jorudas, T. Kaplas, V. Janonis, A. Urbanowicz, I. Kašalynas, Experimental evidence of temperature dependent effective mass in AlGaN/GaN heterostructures observed via THz spectroscopy of 2D plasmons, Applied Physics Letters, 117 (2020) 162101.

[28] Y. Yu, Z. Zheng, H. Qin, J. Sun, Y. Huang, X. Li, Z. Zhang, D. Wu, Y. Cai, B. Zhang, Observation of terahertz plasmon and plasmon-polariton splitting in a grating-coupled AlGaN/GaN heterostructure, Optics Express, 26 (2018) 31794-31807.

[29] X. Cai, A.B. Sushkov, R.J. Suess, M.M. Jadidi, G.S. Jenkins, L.O. Nyakiti, R.L. Myers-Ward, S. Li, J. Yan, D.K. Gaskill, T.E. Murphy, H.D. Drew, M.S. Fuhrer, Sensitive room-temperature terahertz detection via the photothermoelectric effect in graphene, Nature Nanotechnology, 9 (2014) 814-819.

[30] T. Onishi, T. Tanigawa, S. Takigawa, High power terahertz emission from a single gate AlGaN/GaN field effect transistor with periodic Ohmic contacts for plasmon coupling, Applied Physics Letters, 97 (2010) 092117.

[31] V. Jakštas, I. Grigelionis, V. Janonis, G. Valušis, I. Kašalynas, G. Seniutinas, S. Juodkazis, P. Prystawko, M. Leszczyński, Electrically driven terahertz radiation of 2DEG plasmons in AlGaN/GaN structures at 110 K temperature, Applied Physics Letters, 110 (2017) 202101.

[32] V.A. Shalygin, M.D. Moldavskaya, M.Y. Vinnichenko, K.V. Maremyanin, A.A. Artemyev, V.Y. Panevin, L.E. Vorobjev, D.A. Firsov, V.V. Korotyeyev, A.V. Sakharov, E.E. Zavarin, D.S. Arteev, W.V. Lundin, A.F. Tsatsulnikov, S. Suihkonen, C. Kauppinen, Selective terahertz emission due to electrically excited 2D plasmons in AlGaN/GaN heterostructure, Journal of Applied Physics, 126 (2019) 183104.

[33] P. Sai, V.V. Korotyeyev, M. Dub, M. Słowikowski, M. Filipiak, D.B. But, Y. Ivonyak, M. Sakowicz, Y.M. Lyaschuk, S.M. Kukhtaruk, G. Cywiński, W. Knap, Electrical Tuning of Terahertz Plasmonic Crystal Phases, Physical Review X, 13 (2023) 041003.

[34] K. Ahi, Review of GaN-based devices for terahertz operation, Optical Engineering, 56 (2017) 090901.

[35] M. Helm, S. Winnerl, A. Pashkin, J.M. Klopf, J.C. Deinert, S. Kovalev, P. Evtushenko, U. Lehnert, R. Xiang, A. Arnold, A. Wagner, S.M. Schmidt, U. Schramm, T. Cowan, P. Michel, The ELBE infrared and THz facility at Helmholtz-Zentrum Dresden-Rossendorf, The European Physical Journal Plus, 138 (2023) 158.

[36] V.V. Korotyeyev, V.A. Kochelap, V.V. Kaliuzhnyi, A.E. Belyaev, High-frequency conductivity and temperature dependence of electron effective mass in AlGaN/GaN heterostructures, Applied Physics Letters, 120 (2022) 252103.

[37] T. Hofmann, P. Kühne, S. Schöche, J.-T. Chen, U. Forsberg, E. Janzén, N. Ben Sedrine, C. Herzinger, J. Woollam, M. Schubert, Temperature dependent effective mass in AlGaN/GaN high electron mobility transistor structures, Applied Physics Letters, 101 (2012) 192102.

[38] S.A. Mikhailov, Plasma instability and amplification of electromagnetic waves in low-dimensional electron systems, Physical Review B, 58 (1998) 1517.

[39] D.V. Fateev, V.V. Popov, M.S. Shur, Transformation of the plasmon spectrum in a grating-gate transistor structure with spatially modulated two-dimensional electron channel, Semiconductors, 44 (2010) 1406-1413.

[40] Y.M. Lyaschuk, V. Korotyeyev, Interaction of a terahertz electromagnetic wave with the plasmonic system "grating–2D-gas". Analysis of features of the near field, Ukrainian journal of physics, 59 (2014) 495-495.

[41] T.L. Zinenko, A. Matsushima, A.I. Nosich, Surface-plasmon, grating-mode, and slab-mode resonances in the H-and E-polarized THz wave scattering by a graphene strip grating embedded into a dielectric slab, IEEE Journal of Selected Topics in Quantum Electronics, 23 (2017) 1-9.

[42] Y.M. Lyaschuk, S.M. Kukhtaruk, V. Janonis, V.V. Korotyeyev, Modified rigorous coupled-wave analysis for grating-based plasmonic structures with a delta-thin conductive channel: far-and near-field study, JOSA A, 38 (2021) 157-167.

[43] V. Nayyeri, M. Soleimani, O.M. Ramahi, Modeling graphene in the finite-difference time-domain method using a surface boundary condition, IEEE transactions on antennas and propagation, 61 (2013) 4176-4182.

[44] Y. Vahabzadeh, N. Chamanara, C. Caloz, Generalized sheet transition condition FDTD simulation of metasurface, IEEE Transactions on Antennas and Propagation, 66 (2017) 271-280.

[45] V. Korotyeyev, V. Kochelap, S. Danylyuk, L. Varani, Spatial dispersion of the high-frequency conductivity of two-dimensional electron gas subjected to a high electric field: Collisionless case, Applied Physics Letters, 113 (2018).

[46] S. Rudin, Viscous hydrodynamic model of non-linear plasma oscillations in two-dimensional conduction channels and application to the detection of terahertz signals, Optical and quantum electronics, 42 (2011) 793-799.

[47] S. Rudin, G. Rupper, Plasma instability and wave propagation in gate-controlled GaN conduction channels, Japanese Journal of Applied Physics, 52 (2013) 08JN25.

[48] V. Korotyeyev, Theory of high-field electron transport in the heterostructures $Al_xGa_{1-x}As/GaAs/Al_xGa_{1-x}$ with delta-doped barriers. Effect of real-space transfer, Semiconductor physics, quantum electronics & optoelectronics, (2015) 1-11.

[49] K. Yee, Numerical solution of initial boundary value problems involving Maxwell's equations in isotropic media, IEEE Transactions on antennas and propagation, 14 (1966) 302-307.

[50] N. Vermeulen, D. Espinosa, A. Ball, J. Ballato, P. Boucaud, G. Boudebs, C.L. Campos, P. Dragic, A.S. Gomes, M.J. Huttunen, Post-2000 nonlinear optical materials and measurements: data tables and best practices, Journal of Physics: Photonics, 5 (2023) 035001.

[51] M.C. Hoffmann, D. Turchinovich, Semiconductor saturable absorbers for ultrafast terahertz signals, Applied Physics Letters, 96 (2010).

[52] D. Dietze, J. Darmo, K. Unterrainer, THz-driven nonlinear intersubband dynamics in quantum wells, Optics Express, 20 (2012) 23053-23060.

[53] H.R. Seren, J. Zhang, G.R. Keiser, S.J. Maddox, X. Zhao, K. Fan, S.R. Bank, X. Zhang, R.D. Averitt, Nonlinear terahertz devices utilizing semiconducting plasmonic metamaterials, Light: Science & Applications, 5 (2016) e16078-e16078.

[54] B. Lee, C. In, J. Moon, T.H. Kim, S. Oh, T.W. Noh, H. Choi, Terahertz-driven hot Dirac fermion and plasmon dynamics in the bulk-insulating topological insulator $Bi_2Se_3$, Physical Review B, 105 (2022) 045307.

The table of contents entry should be 50–60 words long and should be written in the present tense. The text should be different from the abstract text.

C. Author 2, D. E. F. Author 3, A. B. Corresponding Author\* ((same order as byline))

**Title** ((no stars))

ToC figure ((Please choose one size: 55 mm broad × 50 mm high **or** 110 mm broad × 20 mm high. Please do not use any other dimensions))

# Supporting Information

**Extreme Terahertz Nonlinearity of AlGaN/GaN-based Grating-Gate Plasmonic Crystals**

*Pavlo Sai[1,2], Vadym V. Korotyeyev[1,3], Dmytro B. But[1,2], Maksym Dub[1,2], Dmitriy Yavorskiy[1,2], Jerzy Łusakowski[4], Mateusz Słowikowski[2], Serhii Kukhtaruk[1,3], Yurii Liashchuk[1,3], Jeong Woo Han[5,6], Christoph Böttger[5], Alexej Pashkin[7], Stephan Winnerl[7], Wojciech Knap[1,2], and Martin Mittendorff[5]\**

[1]Institute of High Pressure Physics PAS, 01-142 Warsaw, Poland
[2]CENTERA, CEZAMAT, Warsaw University of Technology, 02-822 Warsaw, Poland
[3]V. Ye. Lashkaryov Institute of Semiconductor Physics (ISP), NASU, 03028 Kyiv, Ukraine
[4]Faculty of Physics, University of Warsaw, ul. Pasteura 5, 02-093 Warsaw, Poland
[5]Universität Duisburg-Essen, Fakultät für Physik, 47057 Duisburg, Germany
[6]now at Department of Physics Education, Chonnam National University, Gwangju 61186, South Korea
[7]Helmholtz-Zentrum Dresden-Rossendorf, 01328 Dresden, Germany
E-mail: martin.mittendorff@uni-due.de

## 1. Study of the temperature dependence of the cyclotron resonance

The cyclotron resonance (CR) measurements were carried out on the unprocessed AlGaN/GaN structure placed in a variable temperature insert at the center of a superconducting coil, with a 2DEG perpendicular to the magnetic field. The sample temperature was varied and controlled by a PID controller between 10 K and 290 K. A monochromatic THz radiation was guided to the sample with a round stainless steel waveguide with the internal diameter of 8 mm, which ended 5 mm from the surface of the sample. The electronic Virginia Diodes, Inc. (VDI) source was used to generate THz radiation with the frequency $f_c = 0.66$ THz. A signal transmitted through the sample was guided through a second waveguide outside of the cryostat and detected with a pyroelectric detector using a lock-in technique. The transmission signal was recorded for each stabilized temperature as a function of the magnetic field up to 10 T. The dependence of the transmitted signal on the magnetic field at different temperatures is presented in **Figure S1**(a). Here, all curves are vertically shifted for clarity. Surprisingly, even at temperatures higher than 200 K, CR remains visible. The resonant values of the magnetic field, $B_c$, corresponding to minima of each curve were used to recalculate the electron effective mass as follows: $m^* = eB_c/2\pi c f_c$. The obtained temperature dependence of the effective mass, $m^*(T)$ is shown in Figure S1(b) by the black dots. The interpolated curve presented by the solid line was used in FDTD simulations.

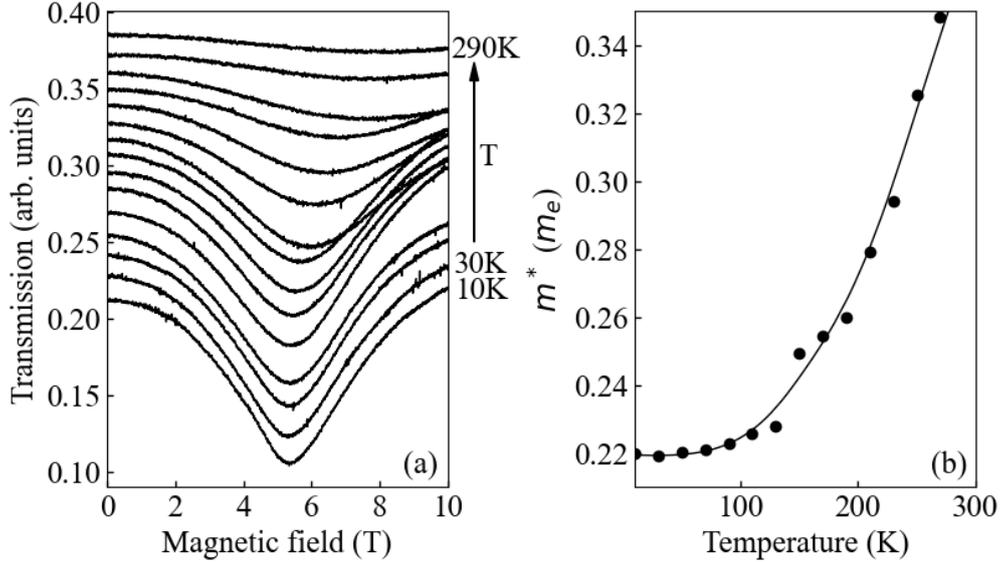

**Figure S1.** Transmission spectra of 0.66 THz beam as a function of the magnetic field at different temperatures (a). Temperature dependence of the effective mass (b).

## 2. Applicability of viscous HD transport model

The viscous HD model presented here, was derived as a perturbative procedure from the quasi-classical Boltzmann transport equation (for details, see [42]) which is formulated for the electron distribution function, $F(\vec{p}, x, t)$, (note vector $\vec{p}$ has two components $p_x$ and $p_y$):

$$\frac{\partial F(\vec{p},x,t)}{\partial t} + \frac{p_x}{m^*}\frac{\partial F(\vec{p},x,t)}{\partial x} - e\left[E^{2D}(x,t) - \frac{\partial \varphi_0}{\partial x}\right]\frac{\partial F(\vec{p},x,t)}{\partial p_x} = \hat{L}_{qel}\{F(\vec{p},x,t)\} + \hat{L}_{pop}\{F(\vec{p},x,t)\} +$$
$$\hat{L}_{e-e}\{F(\vec{p},x,t)\}, \tag{S1}$$

where $\hat{L}_{qel,pop,e-e}\{F(\vec{p}, x, t)\}$ are the scattering integrals for quasi-elastic, electron-LO phonon and electron-electron (*e-e*) scattering mechanisms, respectively. This procedure involves the expansion of distribution functions $F(\vec{p}, x, t) = F^{(0)}(\vec{p}, x, t) + F^{(1)}(\vec{p}, x, t)$ where $F^{(1)}(\vec{p}, x, t)$ is perturbation. Starting with a drifted Fermi-Dirac distribution,

$$F^{(0)} = \left(1 + exp\left[\frac{(p_x - P_D)^2 + p_y^2}{P_T^2} - \xi_F\right]\right)^{-1}, \tag{S2}$$

where $\xi_F$ is the dimensionless Fermi level, $\epsilon_F/k_B T_e$ (note for 2DEG, $n(x,t) = \frac{P_T^2}{2\pi\hbar^2}ln[1+exp\xi_F]$), multiplying both sides of Equation (12) by 1, $p_x$ and $p^2$, and integrating them over all $\vec{p}$, we can come to the Equations (3)-(5) of main text without terms proportional to the viscosity and the thermal conductivity. The latter terms are obtained as a first-order corrections with respect to Knudsen number, using the e-e collision integrals in the form of relaxation time approximation $\hat{L}_{e-e}\{F(\vec{p},x,t)\} \approx -F^{(1)}/\tau_{e-e}$.

Functions $Q_p$ and $Q_E$ in Equations (4) and (5) are the results of averaging the sum of scattering integrals, $\hat{L}_{qel}\{F^0(\vec{p},x,t)\}+ \hat{L}_{pop}\{F^0(\vec{p},x,t)\}$, with weights $p_x$ and $p^2$, respectively.

In general, the applicability of HD model implies that e-e scattering is the fastest scattering mechanisms in 2DEG. It means that e-e scattering rate $\frac{1}{\tau_{e-e}}$ should be much larger than total scattering rate $\frac{1}{\tau_{sc}}$ corresponding other scattering mechanisms. It should be noted that the 2DEG in our unbiased AlGaN/GaN heterostructures with the equilibrium electron concentrations $n_0 \approx 6-8 \times 10^{12}$ cm$^{-2}$ is strongly degenerated in a wide temperature range up to the room temperatures. The calculation of the Fermi level at $n_0 = 7 \times 10^{12}$ cm$^{-2}$ and $m^* = 0.22 \times m_e$ provides the following numbers: $E_F = 75$ meV or $T_F = 875$ K, the Fermi velocity, $V_F = \sqrt{2E_F/m^*} \approx 3.5 \times 10^7$ cm s$^{-1}$. In this case, the $1/\tau_{e-e}$ can be estimated as follows [1]: $\frac{1}{\tau_{e-e}} = \frac{E_F}{\hbar} \times \left(\frac{T_e}{T_F}\right)^2 ln\left(\frac{T_F}{T_e}\right)$.

At the equilibrium temperature, $T_e = T_0 = 10K$, $\frac{1}{\tau_{e-e}} = 0.06$ THz ($\tau_{e-e} = 14$ ps) the corresponding mean free path, $l_{e-e} = V_F \tau_{e_e} = 5.2$ μm. The scattering rate $\frac{1}{\tau_{sc}}$ can be estimated using the low-field mobility, $\mu_0$, $\frac{1}{\tau_{sc}} = \frac{e}{m^*\mu_0}$. At reported value of $\mu_0$=7500 cm$^2$ V$^{-1}$s$^{-1}$, $\frac{1}{\tau_{sc}} = 1.06$ THz ($\tau_{sc} = 0.96$ ps) and $l_{sc} = V_F\tau_{sc} = 0.32$ μm. Thus, we see that at the near-equilibrium conditions the criteria of viscous HD model are not fulfilled and we have rather a ballistic regime of a high-frequency electron dynamics with respect to e-e scattering. Nevertheless, at small-amplitudes of the external signal (neglecting the 2DEG thermal heating, $T_e \approx T_0$, as well as the transient modification of electron concentration profile, $n(x,t) \approx n_0$ and the viscosity effect), HD model provides a transition to the conventional Drude model of the high-frequency response of the 2DEG. This circumstance was pointed out in Ref. [2]. Additionally, we checked that a linear Drude model and discussed HD transport model provide almost the same results in electrodynamic calculations of the transmission spectra for amplitudes of the incident pulse, $E_{max} \approx 0.1$ kV cm$^{-1}$.

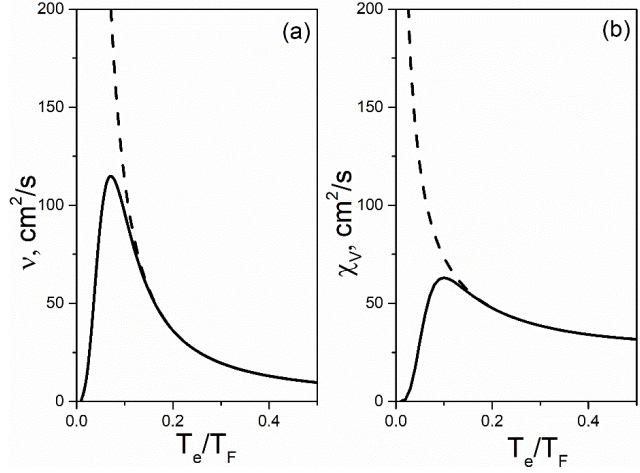

**Figure S2.** Kinematic viscosity (a) and normalized thermal conductivity (b) of the 2DEG in GaN QW as a function of the dimensionless electron temperature. Dashed lines are $\nu(T_e)$ and $\chi_V(T_e)$ given by Equations (6) from the main text. Solid lines are interpolated dependencies $\nu_{int} = \nu(T_e) th\left(\frac{20T_e}{T_F}\right)^6$ and $\chi_{V,int} = \chi(T_e) th\left(\frac{20T_e}{T_F}\right)^6$.

However, under the action of larger-amplitude pulses ($E_{max} \approx 1 - 5$ kV cm$^{-1}$), the 2DEG is subjected to an essential heating, the electron temperature can reach the values of 120-150 K. At this, we have already had a good satisfaction of the HD transport model criteria. So, for $T_e = 150$ K, $\frac{1}{\tau_{e-e}} = 5.9$ THz ($\tau_{e-e} = 0.16$ ps), the corresponding mean free path, $l_{e-e} = V_F \tau_{e_e} = 0.058$ μm. Moreover, the magnitudes of the electric fields acting on 2DEG can be several times larger than the amplitude of the incident pulse due to the near-field effects and excitations of the plasmon oscillations.

In the situation of a ballistic transport regime, the viscosity effects should disappear. As a result, formulas (6) providing the rapid increase of the factors $\nu$ and $\chi_V$ at low temperatures are not applicable. In order to avoid this drawback of the viscous HD model, in calculations, we used interpolated temperature dependences of the viscosity coefficient, $\nu_{int}$ and the thermal conductivity coefficient, $\chi_{V,int}$. Both are shown in **Figure S2.** by solid lines. The interpolated dependencies $\nu_{int}$ and $\chi_{V,int}$ provide the transition of the viscous HD model to model of Drude-Lorentz response of 2DEG for low-power pulses with $E_{max} < 0.1$ kV cm$^{-1}$ and at low equilibrium temperatures $T_0 \sim 10$ K.

## 3. Optical characteristics in the linear regime

The preliminary FDTD simulations of the optical characteristics were performed for the parameters of sample S13. The chosen parameters of the metal grating and dielectric layers were the following: $a_G = 1$ μm, $w_G = 0.5$ μm, $d_g = 0.05$ μm, $\sigma_M = 4 \times 10^{17}$ s$^{-1}$, $d_b = 0.022$ μm, $\epsilon_b = 8.9$, $d_s = 500$ μm, $\epsilon_s = 9.72$ (see **Figure 5**).

**Figure S3** provides the results of FDTD calculations of a transient waveform of the incident and transmitted pulses (a), their power Fourier spectra (b) and spectra of the transmission coefficient (c). Here, we consider the delocalized phase and a uniform lateral concentration profile of the 2DEG with $n_0 = 6.6 \times 10^{12}$ cm$^{-2}$. To calculate the optical characteristics in a wide spectral range, a relatively short incident pulse with the transient width, $\delta t = 0.5$ ps, is assumed. Also, we set $E_{max} = 0.1$ kV cm$^{-1}$ and $\nu_c = 1.84$ THz. The parameters for calculations were listed in the Figure 6 caption.

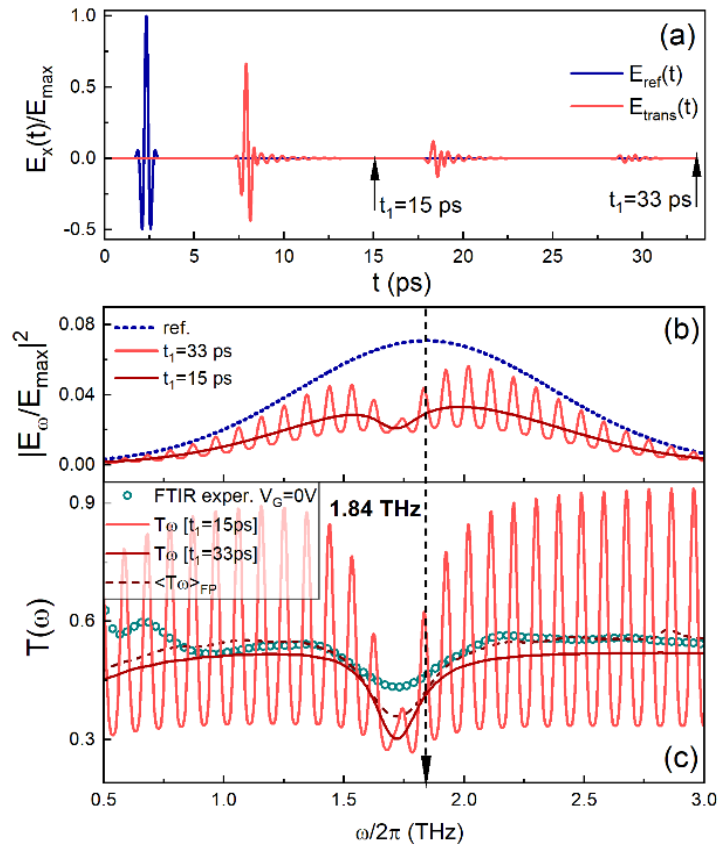

**Figure S3.** The form of the incident and transmitted pulses in the time domain (recorded at the coordinate $z = z_5$, after the bottom interface of the substrate) (a). A normalized power Fourier spectra of the corresponding pulses (b). Spectra of transmission coefficients (c). Calculations according to Equation (7): solid lines correspond to the power Fourier spectra from panel (b), dashed line is the $<T_\omega>_{FP}$ coefficient. Open dots are the experimental FTIR results. The black vertical dashed line denotes the FEL carrier frequency, $\nu_c = 1.84$ THz.

As seen from Figure S3 (a), the transmitted pulse acquires the form with several echoes (at least three of them are well resolved) coming from multiple reflections in the thick substrate. In this case, the spatial half-width of the *em* pulse in the substrate, $\delta z = c\delta t/\sqrt{\epsilon_s} = 48$ μm, is less than substrate thickness, $d_s = 500$ μm. Additionally, the form of the transmitted pulse (the echos) experiences a significant reformation compared to the incident pulse.

The power Fourier spectra of the transmitted pulse, shown in Figure S3 (b) by solid curves, are obtained using two different time windows [0, 33 ps] and [0, 15 ps], respectively. In the former case, the Fourier spectrum has the Gaussian-like form with a dense modulation by the Fabri-Pérot (FP) fringes. Note, that in the spectral band of 2D plasmon resonance, the amplitude of FP oscillations is strongly suppressed. The same effect is observed in the corresponding spectrum of the transmission coefficient (see the curve calculated for $t_1 = 33$ ps in Figure S3 (c)). In the latter case, taking $t_1 = 15$ ps, one can avoid the emergence of the FP fringes. As a result, the power Fourier spectrum, $|E_\omega|^2$, as well as the transmission coefficient, $T_\omega$, have a more regular form with a well-resolved dip of the 2D plasmon resonance. The $T_\omega$, obtained in such a way, describes well the result of the FTIR measurements (see dotted curve). Alternatively, the averaged transmission coefficient, $<T_\omega>_{FP}$, with respect to a period of FP oscillations (see the dashed curve in Figure S3 (c)) can be used for the analysis of the experimental data. As seen, this quantity provides a better fitting of the FTIR transmission spectrum.

The simulations of optical characteristics for the same sample in the localized phase were performed using the equilibrium concentration profile of the 2DEG shown in **Figure S4** (a). In contrast to Ref. [33], where the abrupt step-like profile was considered, here, we used a particular smoothing of $n_0(x)$ at the boundary of gated and ungated parts of 2DEG (the width of transition area of ~20 nm).

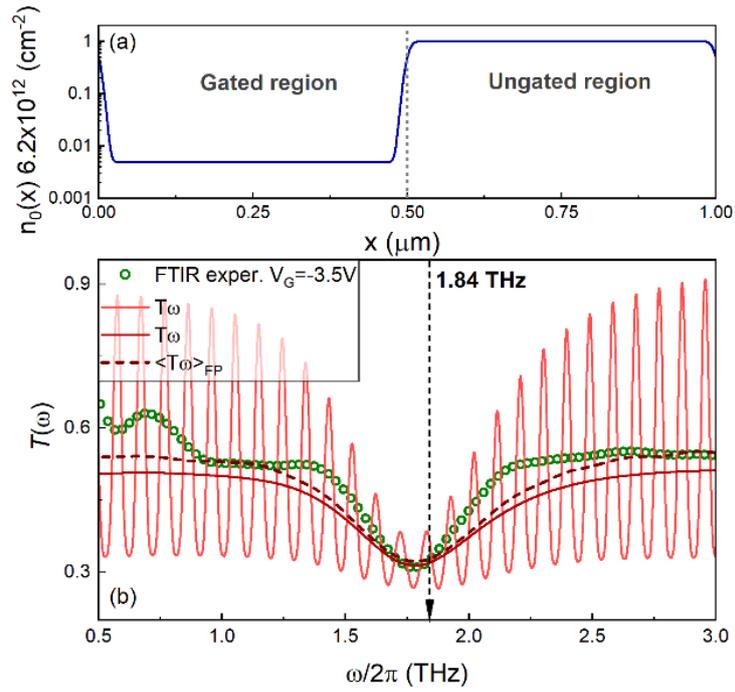

**Figure S4.** The shape of the concentration profile $n_0(x)$ used in calculations (a). The same as in panel (c) of Figure 7 for samples S13 in the localized phase (b).

As seen, the results of FDTD calculations of spectral transmission coefficient spectra (Figure S4 (b)) also well reproduce the results of FTIR THz characterization of sample S13 in the localized phase.